Electronic structure of compressively strained thin film $La_2PrNi_2O_7$

Bai Yang Wang[1,2*], Yong Zhong[1,2], Sebastien Abadi[1,3,4], Yidi Liu[1,3], Yijun Yu[1,2], Xiaoliang Zhang[5], Yi-Ming Wu[6], Ruohan Wang[1,3], Jiarui Li[1,3], Yaoju Tarn[1,2], Eun Kyo Ko[1,2], Vivek Thampy[7], Makoto Hashimoto[7], Donghui Lu[7], Young S. Lee[1,3], Thomas P. Devereaux[1,4,8], Chunjing Jia[5], Harold Y. Hwang[1,2], Zhi-Xun Shen[1,2,3,4*]

[1] *Stanford Institute for Materials and Energy Sciences, SLAC National Accelerator Laboratory, Menlo Park, CA 94025, USA*
[2] *Departments of Applied Physics, Stanford University, Stanford, CA 94305, USA*
[3] *Departments of Physics, Stanford University, Stanford, CA 94305, USA*
[4] *Geballe Laboratory for Advanced Materials, Department of Physics and Applied Physics, Stanford University, Stanford, California 94305, USA*
[5] *Department of Physics, University of Florida, Gainesville, 32611, FL, USA*
[6] *Stanford Institute for Theoretical Physics, Stanford University, Stanford, California 94305, USA*
[7] *Stanford Synchrotron Radiation Lightsource, SLAC National Accelerator Laboratory, Menlo Park, CA 94025, USA*
[8] *Department of Materials Science and Engineering, Stanford University, Stanford, CA 94305, USA*

*Email: bwang87@stanford.edu (B.Y.W.); zxshen@stanford.edu (Z.X.S.)

**Abstract**

The discovery of superconductivity in the bulk nickelates under high pressure is a major advance in physics. The recent observation of superconductivity at ambient pressure in compressively strained bilayer nickelate thin films has now enabled direct characterization of the superconducting phase through angle resolved photoemission spectroscopy (ARPES). Here we present an *in situ* ARPES study of compressively strained $La_2PrNi_2O_7$ films grown by oxide molecular beam epitaxy, and the ozone treated counterparts with an onset $T_c$ of 40 K, supplemented with results from pulsed laser deposition films with similar $T_c$. We resolve a systematic strain-driven electronic band shift with respect to that of bulk crystals, in qualitative agreement with density functional theory (DFT) calculations. However, the strongly renormalized flat $3d_{z^2}$ band shifts a factor of 5-10 smaller than anticipated by DFT. Furthermore, it stays ~70 meV below the Fermi level, contradicting the expectation that superconductivity results from the high density of states of this band at the Fermi level. We also observed a non-trivial $k_z$ dispersion of the cuprate-like $3d_{x^2-y^2}$ band. Combined with results from both X-ray diffraction and DFT, we suggest that the strained films are under ~5 GPa effective pressure, considerably larger than the naïve expectation from the DFT relaxed structure. Finally, the ~70 meV energy position is intriguingly close to the collective mode coupling more prominently seen in thin films, in the energy range of both oxygen related phonons and the maximum of the spin excitation spectrum.

## Introduction

Since the discovery of high transition temperature ($T_c$) superconductivity in cuprates[1], great efforts have been devoted to searching for its electronic and structural analogs within other transition metal oxide families[2–4]. The recent discovery of superconductivity near 80 K in bulk bilayer $La_3Ni_2O_7$ (LNO) and near 30 K in trilayer $La_4Ni_3O_{10}$ under pressure has cemented the high-$T_c$ nature of nickelate superconductivity and comparisons to the cuprates have already provided new perspectives on the essential ingredients[5,6]. Early experimental and theoretical studies of bulk LNO identified a flat band (γ band) with a large density of states just below the Fermi level with dominantly $3d_{z^2}$ orbital character within an otherwise rather two-dimensional (2D) band structure[7–13]. This γ band has been proposed to cross the Fermi level under high pressure, thereby plays a central role in the emergence of superconductivity, in contrast to the dominance of the $3d_{x^2-y^2}$ band in cuprate superconductors[8–17]. Direct experimental verification of this proposal, however, has been extremely challenging due to the need for large hydrostatic pressure for bulk LNO superconductivity.

In this regard, the latest observation of superconductivity in $(La,Pr)_3Ni_2O_7$ thin films at ambient pressure offers a direct approach for investigating the mechanism of superconductivity[18–20]. This stabilization of the superconducting phase is achieved through a combination of a large compressive epitaxial strain and a fine-tuned post-growth oxidation treatment. In particular, the compressive strain is predicted to induce a systematic difference in the electronic structure between bulk LNO and the thin films. For example, the aforementioned γ band is anticipated to shift further from the Fermi level, by the amount as large as 350 meV with 2.5% strain[21–23]. From a technical point of view, the atomically flat surfaces of thin films make them ideal for direct electronic structure measurements using angle resolved photoemission spectroscopy (ARPES). Indeed, two ARPES studies of compressively strained $(La,Pr)_3Ni_2O_7$ thin films have been reported recently[24,25]. Utilizing both synchrotron and laser light sources with photon energies of 200 eV and 7 eV, the authors highlight a Fermi level crossing of the γ band and a nodeless anomalous gap on the other $3d_{x^2-y^2}$ bands, respectively. However, the reported γ band characteristics differ from both bulk samples and theoretical expectations considerably[7,21–23]. The nature of the nodeless anomalous gap that persists above the superconducting $T_c$ is also under debate[25]. Given the level of discrepancy, further experimental study of the electronic structure is clearly needed.

In this work, we synthesized compressively strained superconducting $La_2PrNi_2O_7$ (LPNO) films of mainly 4-formula-unit-cell thickness utilizing an oxide molecular beam epitaxy (MBE) system, and characterized their electronic structure using an *in situ* connected synchrotron ARPES end station. To eliminate synthesis-recipe-specific artifacts, we also examined *ex situ* transferred thin films of the same composition grown by pulsed laser deposition (PLD) with similar thickness. These films are characterized both before and after an *in situ* ozone treatment, which reduces the oxygen vacancies and stabilizes the superconducting phase with an onset $T_c$ around 40 K. Our photon-energy-dependent ARPES measurements with improved spectral quality resolve a systematic band shift compared with the results from bulk LNO. The downward shift of the strongly renormalized γ band is qualitatively consistent with the theoretical predictions but quantitatively off by a factor of 5-10[21–23]. Specifically, in ozone treated superconducting samples, we find that the γ band remains ~70 meV below the Fermi level, rather than crossing it or shifting downward by the predicted 350 meV[21]. This ~70 meV energy position is intriguingly close to the collective mode coupling more prominently seen in thin films, in the energy range of both oxygen related phonons and the maximum of spin excitation spectrum[7,25–29], hinting the importance of combined effects of strong correlation and lattice dynamics. We also observe a notable coupling along the $c$-axis manifested through a clear $k_z$ dispersion of a $3d_{x^2-y^2}$ band, in contrast to the highly 2D electronic structure found in density functional theory (DFT) calculations[13]. These results lay a key foundation for refining quantitative theoretical models of the superconductivity in the bilayer nickelates.

## Thin film MBE synthesis

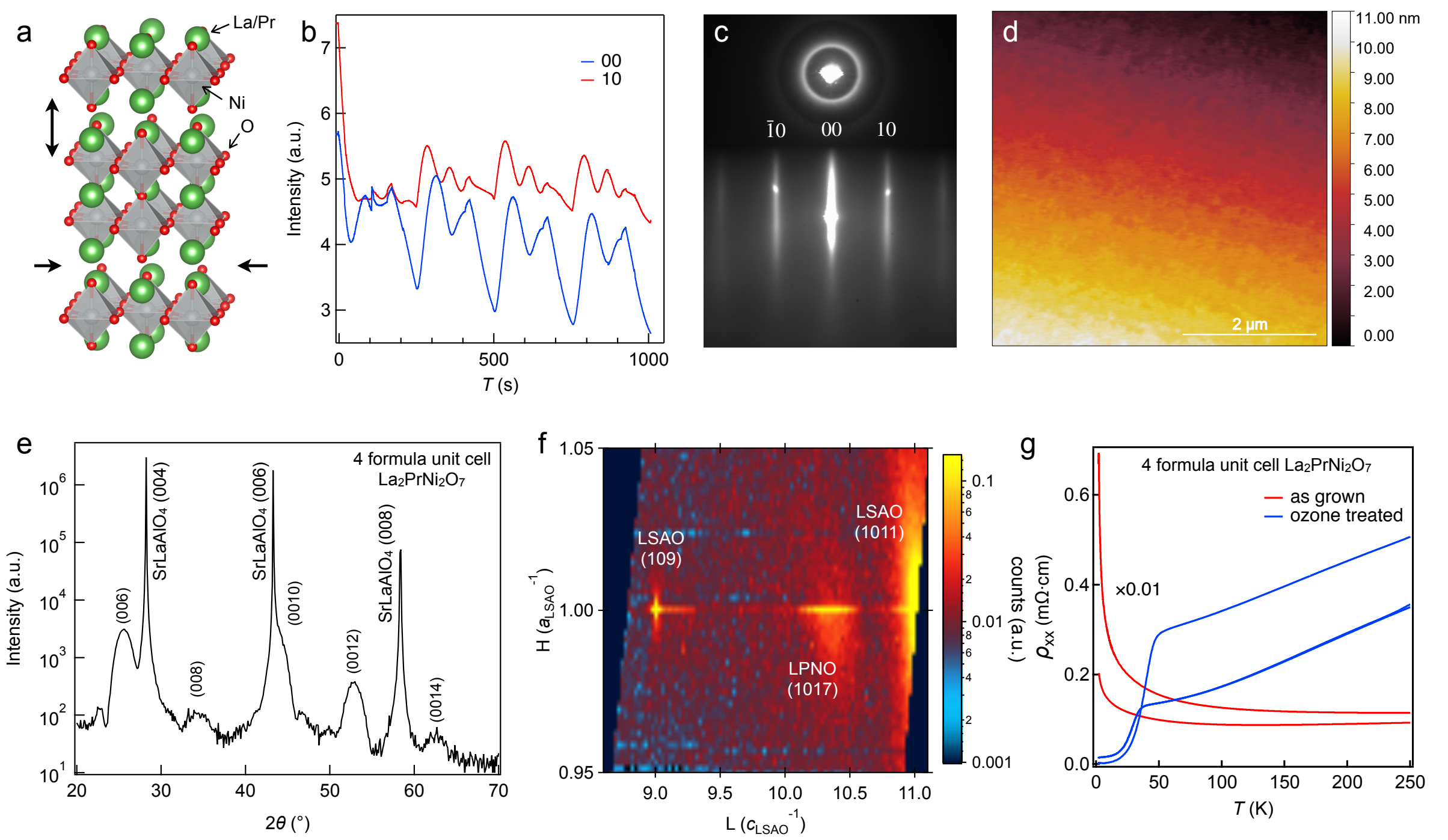

**FIG. 1.** MBE synthesis of compressively strained LPNO thin film. (a) The structural schematic of LPNO with the arrows indicating the epitaxial compressive strain as well as the resulting elongation along the $c$-axis. (b) and (c) Representative reflection high-energy electron diffraction (RHEED) oscillations and pattern of LPNO growth on the SLAO substrate. The blue and red curves correspond to intensity oscillations of the 00 and 10 peaks in the RHEED pattern. (d) Representative atomic force microscopy image of the LPNO film. The root-mean-square roughness is 161.7 pm. (e) X-ray diffraction (XRD) along the $c$-axis for a 4-formula-unit-cell thick LPNO film with the substrate and thin film peaks labeled. (f) Reciprocal space map (RSM) of the same thin film sample show in (e) with the substrate and thin film peaks labeled. (g) Resistivity-temperature curves for two representative as grown (down scaled by a factor of 100) and two ozone treated 4-formula-unit-cell thick LPNO films, synthesized by MBE.

The MBE grown LPNO bilayer nickelate films in this study are deposited on (001)-oriented $SrLaAlO_4$ (SLAO) substrates. Figure 1a shows the crystal structure of the LPNO with the arrows indicating the epitaxial compressive strain as well as the resulting elongation along the $c$-axis. In our MBE growth, the SLAO substrate is first annealed in $5\times10^{-6}$ Torr ozone at 750 °C for 10 minutes. LPNO thin films are then deposited layer by layer at 680 °C in $1.8\times10^{-6}$ Torr ozone in a shutter-controlled mode. During growth, reflection high-energy electron diffraction with a 15 keV electron beam was employed to monitor the growth process and film quality. A representative RHEED oscillation curve and the corresponding surface RHEED pattern are shown in Fig. 1b and 1c. After growth, the sample is first cooled down at 80 °C/min till below 80 °C while maintaining ozone pressure and then transferred into the *in situ* connected synchrotron ARPES end station's sample stage (~7 K) within approximately 20 minutes. The precise control of stoichiometry (La+Pr:Ni ratio) and the monolayer dosage needed for the synthesis of high-quality bilayer nickelate is achieved through the calibration growth of $La_{0.67}Pr_{0.33}NiO_3$ using the co-deposition growth mode[30,31]. In the rest of this paper, the film thickness is 4 formula unit cell unless otherwise specified. The synthesis methods of the PLD grown samples has been previously reported[20]. These thin films are exposed to ambient environment and *ex situ* transferred into the load lock chamber (~$1\times10^{-9}$ Torr) of the ARPES end station in roughly 30 minutes upon cool down.

The as grown MBE thin film shows a sharp top surface with step terraces and root-mean-square roughness of 161.7 pm, as shown in Fig. 1d. The structural quality is characterized by XRD, shown in Fig. 1e, which displays the clear multiple (00l) peaks indexed to the bilayer phase, along with the SLAO substrate peaks. The extracted *c*-lattice constant is 20.72 Å, ~1% larger compared to the bulk values. The ~2% compressive epitaxial strain of the film sample is also confirmed by synchrotron X-ray reciprocal space mapping (RSM) measurement, shown in Fig. 1f. The representative transport measurements also show an insulating temperature dependence at low temperatures (Fig. 1g), which is consistent with previous reports without the additional ozone treatment[18,30,31].

To fill in the oxygen vacancies, both MBE and PLD grown thin films are transferred to an *in situ* connected ozone treatment chamber where they are treated under 900 mTorr of flowing ozone at temperatures within the range of 210 - 330 °C for 30 minutes. Upon completion, the samples are transferred into the *in situ* connected synchrotron ARPES end station sample stage (7 K) in approximately 20 minutes. Within 1 hour after the ARPES measurement, the transport properties of these ozone treated samples are measured *ex situ*, with Fig. 1g showing the superconducting transition of two representative MBE samples with onset $T_c$ comparable to the literature[18–20]. These structural and transport characterizations demonstrate the high quality of the thin film samples in our study.

**Electronic band structure**

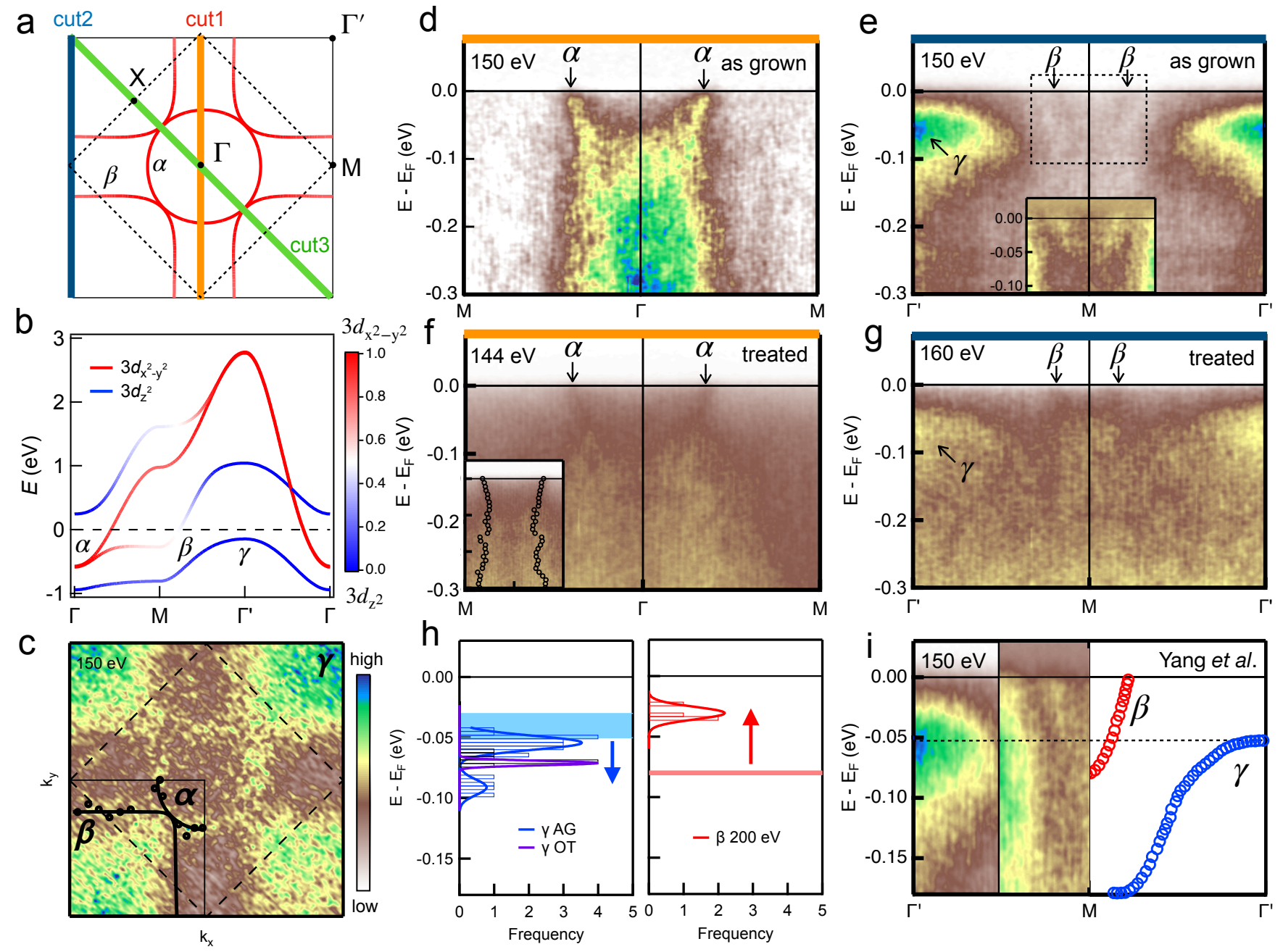

**FIG. 2.** ARPES characterization of the electronic structure. (a) and (b) Schematic Fermi surface map and two-orbital band structure taken from literature[8,21]. The three solid lines in orange, green, and blue colors in (a) mark the directions of the high-symmetry cuts shown in the rest of the figures. The color bar in (b) describes the projection weight of $3d_{x^2-y^2}$ (red) and $3d_{z^2}$ (blue) orbitals. (c) Fermi surface map of an as grown MBE sample, using linear horizontally polarized (LH) photons with an energy of 150 eV. The black circles in the lower left quadrant mark the positions of fitted Lorentzian peaks in the momentum distribution curves (MDC) with the solid black lines as guides to the eye. (d) and (e) Spectra of the same sample in (c) along cut 1 and 2 defined in (a). The inset in (e) shows the region near the Fermi level at the M point. (f) and (g) Spectra of a superconducting PLD sample with ~40 K $T_c$ onset, along cuts 1 and 2. The photon energies used are indicated in the top left corner. The black circles in the inset mark the positions of fitted Lorentzian peaks in the MDC describing the α band. (h) Histogram of the energy position of the γ band top (blue) and β band bottom (red) in all measured thin film samples. There is no obvious distinction in the distribution between the two synthesis methods. Here the β band is measured by 200 eV photons and the γ band mostly by 150 eV photons, except for two samples measured with 144 eV and 160 eV photons. Negligible photon energy dependence for the γ band is observed among these three photon energies. The blue and purple lines are gaussian peak fits of the γ band top distribution for the as grown (AG) and ozone treated (OT) samples, respectively, centering at -54 meV, -91 meV, and -71 meV. The red line centered at -31 meV is similarly plotted for the β band. The light blue shaded region and the light red line indicate the range of the γ band and β band energy positions reported in bulk LNO samples respectively[7]. The blue and red arrows indicating shift directions are guides to the eye. (i) Spectra comparison between the measured band structure of the as grown thin film (left half) and the bulk (right half)[7] along cut 3. The two regions near the Γ and M points of the same spectra in the left half are plotted in two different color scales in order to show the β and γ band positions more clearly.

Figure 2 shows the electronic band structure of the LPNO thin film measured at 7 K. The Fermi surface schematic and the corresponding band dispersions of a relaxed bulk measurement are taken from the literature and presented in Fig. 2a and 2b as a reference, with the two components of the Fermi surface corresponding to the α and β bands with majority $3d_{x^2-y^2}$ orbital character[8,21]. The flat $3d_{z^2}$ band just below the Fermi level is referred to as the γ band. The measured Fermi surface of an as grown MBE film is shown in Fig. 2c with an integration window of 30 meV. It clearly shows the α pocket and a weak presence of the β pocket, which is overshadowed by the γ band positioned at the Γ' point as its large spectral weight leaks into the Fermi surface map.

Two high-symmetry cuts are shown in Fig. 2d-e with their corresponding directions marked in Fig. 2a as solid colored lines. The inset shows a zoomed-in view of the β band dispersion. Together, these two cuts provide a complete illustration of the multiorbital electronic structure of the as grown strained LPNO thin film, where both α and β bands cross the Fermi level while the γ band is located just below. Compared to the bulk values, the top of the γ band shifts to a deeper energy, while the bottom of the β band shifts upward. Upon ozone treatment, the electronic structure remains largely unchanged with only subtle band shifts. Fig. 2f and 2g show two high-symmetry cuts of a representative samples grown by PLD. Superconducting transition with ~40 K onset is confirmed for this samples post-ARPES measurement. Notably, the γ band is resolved to be ~70 meV below the Fermi level, in contrast to the previous report[24]. At the same time, in both as grown (Fig. 2d) and superconducting (Fig. 2f) samples, there is a sign of band renormalization kink of the α band dispersion also around 70 meV below the Fermi level, reminiscent of collective mode coupling at this energy. This is observed in another report as well, with the effect appearing to be more prominent in superconducting films, an aspect that requires more systematic investigation[7,25,28]. This fits into the energy range of oxygen related phonons and matches the maximum of spin excitation spectrum seen in bulk samples[26,27]. These observations could shed light on what role each electronic band plays in forming the superconducting phase.

To provide a statistical comparison between the relaxed bulk and the compressed thin film, Fig. 2h summarizes, in two histograms, the distribution of the γ band top and the β band bottom positions in the measured thin film samples (both as grown and ozone treated), with the range of the bulk values indicated by light blue and red shading[7]. In both panels, the separations between the bulk and thin film values are apparent. Here, the spread in the γ band position is likely due to oxygen content variation stemming from the synthesis and cooling process, which is much reduced for the ozone treated samples. To provide a more direct visual comparison, Fig. 2i juxtaposes the bulk β and γ band dispersions, extracted from ref.[7], with our spectra shown in Fig. 2e along the same cut direction. Here the color scales for the two regions near β and γ bands are adjusted separately to enhance band visibility. Again, a clear and systematic difference between the thin films and the bulk samples is seen. These results are qualitatively consistent with theoretical calculations considering a compressive in-plane strain, although the calculations do overestimate the γ band shift by a large factor of about 5 to 10[21–23].

In Table I, we summarize this discrepancy between experimental observations and theoretical calculations. While there is a quantitative agreement on the β band shift, the calculated γ band shift is significantly larger than experimental observations by an order of magnitude. At a first glance, this may be attributed to an overestimation of the *c*-lattice expansion by the DFT calculation assuming structural relaxation, >2% versus 1% observed. However, even after constraining all lattice constants to experimental values with an effective ~5 GPa compressive pressure, this large quantitative discrepancy regarding γ band shift persists (210 meV *vs.* ~20 meV), suggesting that there are fundamental missing pieces in the current DFT calculations (see Supplementary Information). By renormalizing the calculated bands of a relaxed bulk with a factor to match the measured dispersions, we also obtain the mass renormalization factors for different bands, as listed in Table I[7]. Similar to previous literature on bulk crystals, the 3$d_{x^2-y^2}$ derived bands exhibit relatively weak band renormalizations (~2-4) while the 3$d_{z^2}$ derived γ band shows a strong renormalization (~7)[7].

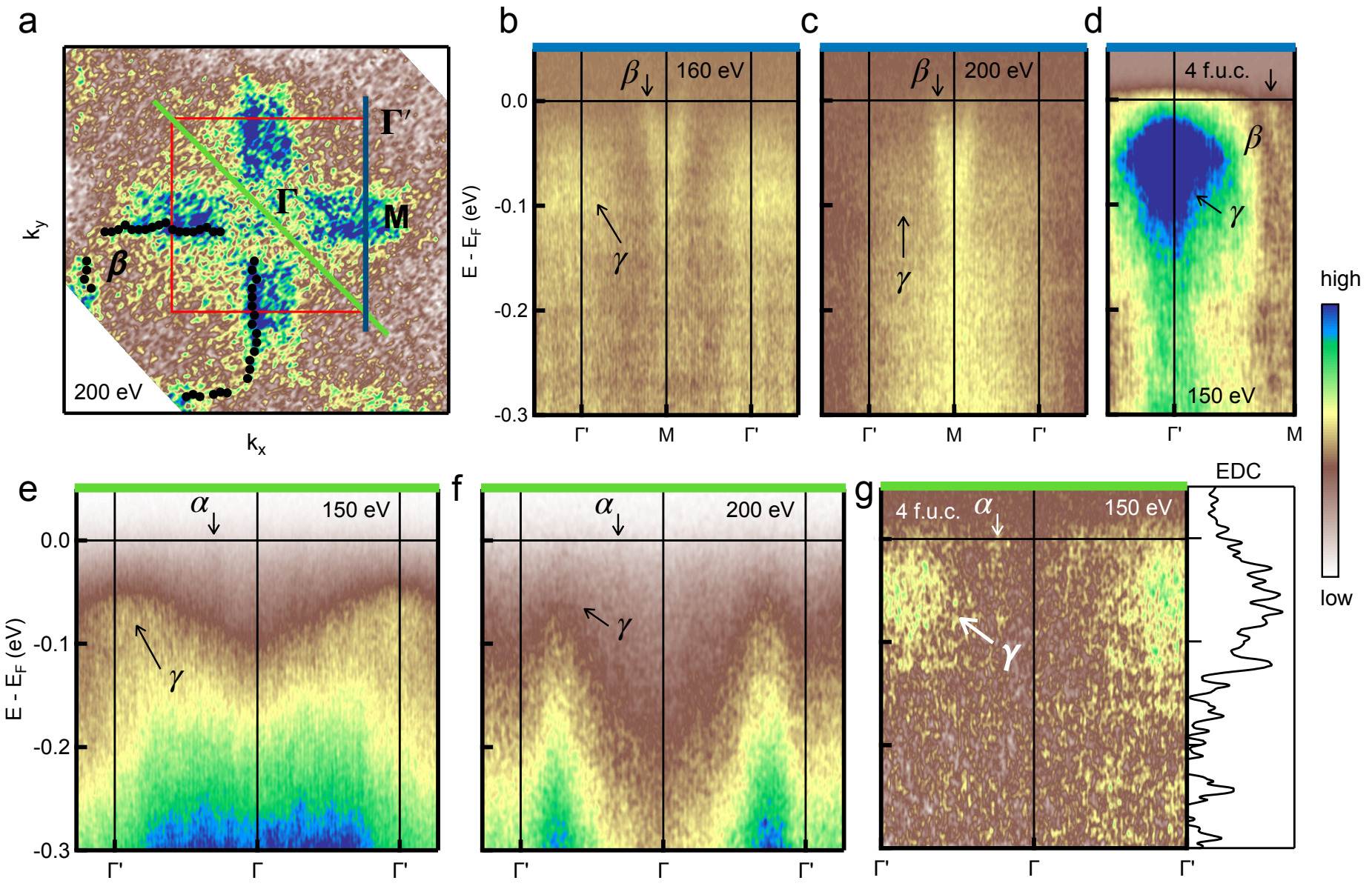

**FIG. 3.** ARPES characterization of a 2-formula-unit-cell thick thin film and a 4-formula-unit-cell thick thin film. (a) Fermi surface map of a 2-formula-unit-cell thick sample measured using linear horizontally polarized (LH) photons with an energy of 200 eV. The black circles in the lower left quadrant mark the positions of fitted Lorentzian peaks in the MDC describing the β pocket. (b) and (c) Spectra of the same sample along cut 2 defined in Fig. 2a. The measurement photon energies are labelled in the top right corner. (d) Spectra of the 4-formula-unit-cell thick sample along the same cut. (e) and (f) Spectra of the 2-formula-unit-cell thick sample along cut 3. (g) Spectra of a superconducting MBE sample plotted in Fig. 1g along cut 3 measured with 150 eV photon. The side panel plots an energy distribution curve (EDC) taken near Γ'.

Here we discuss the measurement of the important γ band position in more detail, which is a non-trivial task complicated by a strong matrix element effect. First, to control for the thickness difference in our comparison to the literature, we also synthesized and characterized a 2-formula-unit-cell thick LPNO thin film. However, we do note that our Pr content differs from the previous report considerably, which may contribute to differences in our observations[24]. Figure 3a shows the Fermi surface of the 2-formula-unit-cell thick sample and Fig. 3b-g show the corresponding spectra taken along two high symmetry directions (marked with solid green and blue lines in Fig. 3a) of this sample and its 4-formula-unit-cell thick counterparts. We observe a substantial photon energy dependent matrix element effect at play. At 200 eV, the resolved Fermi surface and spectra cuts agree with the existing literature, with the intensity of the γ band diminishing at energies close to the Fermi level[24]. This is especially evident in Fig. 3f, making it appealing to extrapolate the γ band dispersion to crossing the Fermi level[24]. However, at 150 eV and 160 eV, the γ band top is clearly resolved at ~65 meV below the Fermi level, as shown in Fig. 3b and 3e. This is not a manifestation of dramatic $k_z$ dispersion of the γ band either, as the γ band is also resolved in Fig. 3c to be below the Fermi level. Furthermore, the almost vertical spectral feature seen in Fig. 3f and the flat γ band top seen in Fig. 3e can be observed simultaneously at 150 eV for thicker samples, as shown in Fig. 3d (4 formula unit cell thick). In this sense, to avoid complications in the interpretation, thicker samples are better suited for future studies. The γ band energy position is also found ~ 70 meV below the Fermi level in the ozone treated superconducting MBE films, with a representative sample shown in Fig. 3g.

**Photon energy dependence**

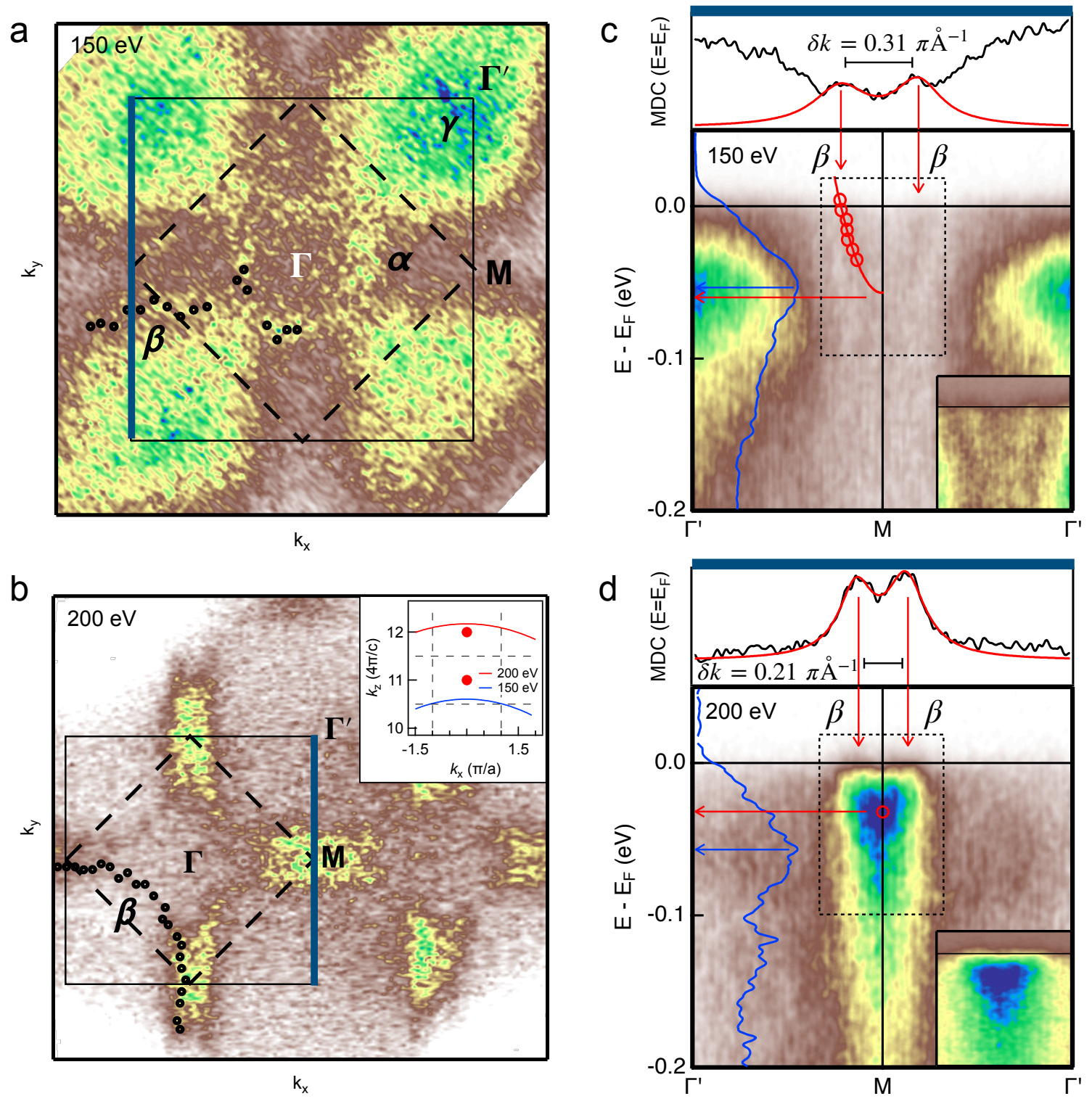

**FIG. 4.** Photon energy dependence of the electronic structure. (a) and (b) Fermi surface maps measured with 150 eV and 200 eV photons, respectively. The inset in (b) shows the $k_z$ - $k_x$ relations of cuts along $k_y = 0$ at photon energies of 150 eV and 200 eV. The black circles in the lower left quadrant mark the positions of fitted Lorentzian peaks in the MDC. (c) and (d) Spectra along cut 2 direction measured with 150 eV and 200 eV photon energy, respectively. The additional top panels show the corresponding MDC taken at Fermi level. The red lines plot two Lorentzian peak fits, marking the $k_F$ positions of the β band dispersion, with their separation in the momentum space labelled. On the left of the plots, EDC taken at Γ' are shown in blue. The peaks are marked with arrows, indicating the energy position of the γ band top. In (c), the β band bottom is determined from a parabolic fit of the MDC-derived β band dispersion. In (d), the β band bottom

is determined from the peak position of an EDC taken at M. The insets show regions near the Fermi level at the M point.

Despite expectations of a strongly 2D electronic structure based on DFT calculations[13], we observe a clear photon energy dependence in the ARPES spectrum. Figure 4 compares the Fermi surface map and a high-symmetry cut at photon energies of 150 eV and 200 eV of the MBE sample reported in Fig. 2c-e. Since as grown and ozone treated samples show similar electronic structure and photon energy dependence, we focus on as grown samples below as its gives better spectral quality. As shown in Fig. 4a and 4b, the β pocket expands considerably as the photon energy increases from 150 eV to 200 eV. In contrast, the presence of the α pocket, as well as the spectral weight from the hidden γ band, is diminished at 200 eV due to matrix element effects. In Fig. 4c and 4d, the two cuts with their corresponding directions marked in Fig. 4a and 4b provide a clearer view of the relative bands shifts and spectral weight changes. At 200 eV, although the intensity of the γ band fades dramatically, its momentum and energy position can still be resolved which remain mostly unchanged compared to the 150 eV cut. Meanwhile, the β band shifts considerably upward. At 150 eV, the γ band top is closer to the Fermi level than the β band bottom, while at 200 eV, the situation is reversed. As a result, the Fermi momentum of the β band decreases by one-third, as shown in the MDCs taken at Fermi level and overlaid on each corresponding cut. These shifts indicate a nontrivial $k_z$ dispersion of the β band, as the 150 eV and 200 eV photon energies correspond to cuts near $k_z = \pi$ and $k_z = 0$, respectively (shown in the inset of Fig. 4b).

Before we discuss in more details the implications of the observed $k_z$ dispersion, an alternative interpretation of the data should be addressed. Epitaxial thin films undergo a strain relaxation as its thickness exceeds a relaxation length. In this sense, the photon energy dependence may be a manifestation of strain relaxation as the 200 eV photons probe deeper into the more highly strained section of the film. However, in terms of experimental evidence, we observe no relaxation in our RSM measurement and the probing depth difference between the two energies is negligible compared to the *c*-lattice constant. Theoretical studies also find the γ band to be much more sensitively tuned by the compressive in-plane strain which is opposite to our observation[21–23]. In this sense, the photon energy dependence indeed suggests a non-trivial hopping integral along the *c*-axis. Furthermore, we call attention to the orbital projection weight shown in Fig. 2b, where the β band has a strong $3d_{x^2-y^2}$ orbital character and the γ band has predominantly a $3d_{z^2}$ orbital character. While it is certainly counter-intuitive for the $3d_{x^2-y^2}$ orbital to exhibit a larger $k_z$ dispersion, such a three dimensional (3D) hybridization could result from the substantial Ni-planar O buckling reported recently (see Supplementary Information)[32]. In addition, we find that this $k_z$ dispersion combined with the overall electronic structure cannot be satisfactorily described by a simple tight-binding model, likely a manifestation of strong correlation (see Supplementary Information). These observations highlight the important role of crystal structure in bilayer nickelates and suggest the need for theoretical models that consider strong correlation effects as well as the 3D nature of the electronic structure.

## Conclusion

In this work, we report the synthesis of high quality compressively strained superconducting LPNO thin films using an ozone-assisted oxide MBE, along with ARPES characterization of their basic electronic structures at an *in situ* connected synchrotron end station. We observe a clear and systematic difference in the electronic structure between compressively strained LPNO thin films and bulk LNO, which qualitatively agrees with previous theoretical predictions but quantitively differs by a factor of 5-10 regarding the $\gamma$ band position. Specifically, instead of a 210 meV downward shift, we resolve the $\gamma$ band top at around 70 meV below the Fermi level for both as grown and ozone treated films, despite an effective ~5 GPa volume compression from XRD and DFT results. We highlight that this observation of coincidence between the flat $\gamma$ band and mode coupling energies, on one hand, contradicts the expectation that the flat band crosses the Fermi level and thereby drives the superconductivity. On the other hand, it fits into the energy range of both oxygen related phonons and experimentally observed spin excitations. It indicates the non-trivial interplay and importance of strong correlation and lattice dynamics in these strained films, where LPNO is under much higher effective pressure than DFT results. Additionally, we observe a distinct $3d_{x^2-y^2}$ orbital-specific $k_z$ dispersion in our photon energy dependence study, indicative of a non-trivial 3D character of the $3d_{x^2-y^2}$ orbitals enabled by a buckled Ni-planar O bond angle. These findings provide insights into what role each electronic band plays in stabilizing the superconductivity and suggest the need for further theoretical studies that consider the 3D nature of the electronic structure.

## Methods

### Sample preparation and XRD characterization

For both MBE and PLD grown LPNO thin films, the $SrLaAlO_4$(001) substrates (MTI Corporation) were sonicated in acetone and IPA. The PLD grown LPNO thin films were deposited with a KrF excimer laser (wavelength 248 nm). During the growth, the substrate temperature was maintained at 660 °C, under oxygen partial pressure $p(O_2)$ of 150 mTorr. The pulsed laser repetition rate during the deposition was 5 Hz. The laser fluence was 0.56 J/cm$^2$ with a spot size of 3.3 mm$^2$. After growth, XRD data were measured using a monochromated Cu K$\alpha$1 source ($\lambda$ = 1.5406 Å). Additional XRD and RSM measurement was performed at beamline BL17-2 of Stanford Synchrotron Radiation Lightsource, using a photon energy of 8.33 keV at room temperature. Diffraction signals were collected using a PILATUS3 100K-M detector.

### Transport measurements

The thin film samples were contacted with aluminum wires by an ultrasonic wire bonder and measured in 4He cryostats (base temperature ~1.8 K). The 4-terminal measurements were carried out with 10 μA current.

### ARPES measurements

The ARPES measurements were performed at beamline BL5-2 of Stanford Synchrotron Radiation Lightsource with a Scienta DA30 analyzer. The energy resolution is 26 meV at 200 eV photon energy and better at lower photon energies. The base pressure during the ARPES measurements was better than $3 \times 10^{-11}$ Torr and the measurement temperature of the sample stage was maintained at 7 K.

### Data Availability

The data that support the findings of this study are available from the corresponding authors upon reasonable request.

### Acknowledgement

We thank Chun Lin for helpful discussions. This work was supported by the US Department of Energy (DOE), Office of Science, Office of Basic Energy Sciences, Materials Sciences and Engineering Division, under contract DE-AC02-76SF00515. The work at Advanced Light Source, Lawrence Berkeley National Laboratory, was supported by US DOE under contract DE-AC02-05CH11231.

### Author contributions

B.Y.W. and Z.X.S. conceived and designed the project. B.Y.W. synthesized the MBE grown films with help from Y.Z. and R.W.. Y.L. and E.K.K. synthesized the PLD grown films with the supervision from H.Y.H.. B.Y.W., Y.L., Y.Z., Y.Y., J.L., V.T., R.W., and Y.T. performed the XRD and transport characterizations with the supervision from Y.S.L. and H.Y.H.. B.Y.W. and Y.Y. performed the ozone annealing study. B.Y.W. and S.A. performed the ARPES measurement with help from Y.Z., R.W., M.H., and D.L.. X.Z., B.Y.W., Y.W., T.P.D., and C.J. performed the tight-binding model calculation. B.Y.W. and Z.X.S. analyzed the data and wrote the manuscript with input from all authors.

**Competing Interest**
The authors claim no competing interest.

Supplementary Information is available for this paper.

Correspondence and requests for materials should be addressed to Zhi-Xun Shen, Bai Yang Wang.

| | Band position (meV) (α, β: band bottom; γ: band top) | | | | *c*-lattice constant (Å) | | | Renormalization factor | |
|---|---|---|---|---|---|---|---|---|---|
| Band | Relaxed bulk | LPNO thin film | Relaxed DFT | 2.5% compressed DFT | 2.5% compressed DFT w/ relaxation | LPNO thin film | Relaxed bulk | Relaxed bulk | LPNO thin film |
| α | -248 | N/A | -750 | -650 | 21.09 | 20.72 | 20.52 | 1.8/2.3 | 2.9±0.9 |
| β | -80 | -31(200eV) /-65(150eV) | -203 | -187 | | | | 2.6/2.3 | 3.8±0.9 |
| γ | -50 | -54/-91/-71 | -78 | -406 | | | | 8.3/5.5 | 7±3 |

**Table I.** Energy positions and renormalization factors of the α, β, and γ bands and *c*-lattice expansion. The energy positions refer to the band bottom for the α and β bands, and band top for the γ band. For thin film samples, the two energy positions of the β band listed correspond to measurements using 150 eV and 200 eV photons. The γ band positions listed correspond to the center positions of the two gaussian fits of the as grown samples and one gaussian fit of the ozone treated samples shown in Fig. 2h. The DFT calculated *c*-lattice expansion corresponds to a 2.5% in-plane biaxial compressed and otherwise freely relaxed structure calculation. The experimental band positions and renormalization factors for the bulk are extracted from ref.[7] and the DFT calculated values are taken from ref.[21]. The errors in the renormalization factor for the thin films reflect the uncertainty in determining the band renormalization factors. Note that the band renormalization kink in the α band prevents us from extracting its band bottom position and the lack of dispersion in the γ band also makes the estimation of its renormalization factor difficult.

Supplementary Information for "Electronic structure of compressively strained thin film $La_2PrNi_2O_7$"

Bai Yang Wang[1,2*], Yong Zhong[1,2], Sebastien Abadi[1,3,4], Yidi Liu[1,3], Yijun Yu[1,2], Xiaoliang Zhang[5], Yiming Wu[6], Ruohan Wang[1,2], Jiarui Li[1,3], Yaoju Tarn[1,2], Eun Kyo Ko[1,2], Vivek Thampy[7], Makoto Hashimoto[7], Donghui Lu[7], Young Lee[1,3], Thomas P. Devereaux[1,4,8], Chunjing Jia[5], Harold Y. Hwang[1,2], Zhi-Xun Shen[1,2,3,4*]

*[1] Stanford Institute for Materials and Energy Sciences, SLAC National Accelerator Laboratory, Menlo Park, CA 94025, USA*
*[2] Departments of Applied Physics, Stanford University, Stanford, CA 94305, USA*
*[3] Departments of Physics, Stanford University, Stanford, CA 94305, USA*
*[4] Geballe Laboratory for Advanced Materials, Department of Physics and Applied Physics, Stanford University, Stanford, California 94305, USA*
*[5] Department of Physics, University of Florida, Gainesville, 32611, FL, USA*
*[6] Stanford Institute for Theoretical Physics, Stanford University, Stanford, California 94305, USA*
*[7] Stanford Synchrotron Radiation Lightsource, SLAC National Accelerator Laboratory, Menlo Park, CA 94025, USA*
*[8] Department of Materials Science and Engineering, Stanford University, Stanford, CA 94305, USA*

*Email: bwang87@stanford.edu; zxshen@stanford.edu

**First-principles density functional theory calculations**

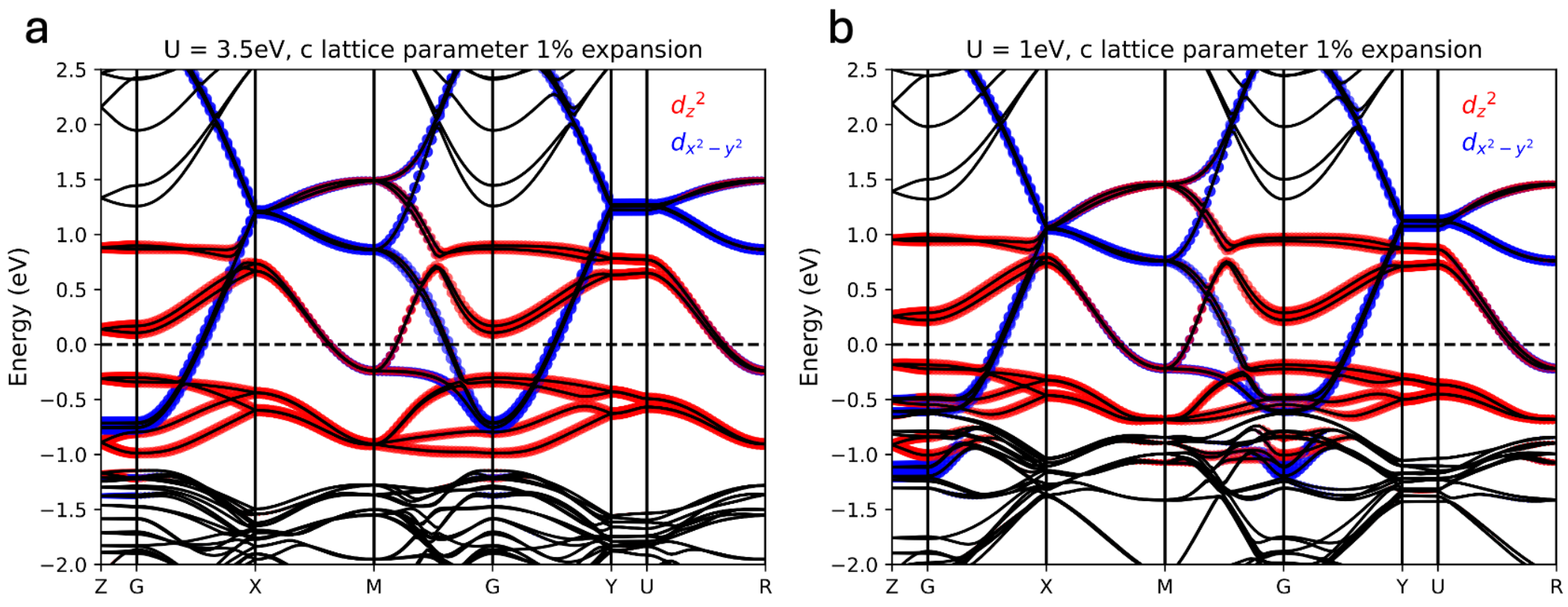

**FIG. S1.** Orbital-resolved band structure of bilayer nickelates $La_3Ni_2O_7$ under 2% in-plane compressive strain using DFT+$U$ calculations. The calculations are performed with the $c$ lattice parameter fixed to the experimental value, 1% larger than the unstrained case, and with (a) $U$ = 3.5 eV and (b) $U$ = 1 eV. Blue and red show the orbital contents from Ni $d_{x2\text{-}y2}$ and $d_{z2}$ orbitals.

To investigate the electronic properties of non-magnetic $La_3Ni_2O_7$ under strain, we perform first-principles density functional theory (DFT) calculations. The lattice parameters and atomic positions of the unstrained structures were fully relaxed, using the Vienna ab-initio Simulation Package (VASP)[1] within the DFT+$U$ framework, with $U = 3.5$eV. We adopt lattice parameters according to the experimental conditions, where $a$ and $b$ shrink by 2% while $c$ expands by 1%. The atomic positions are fully relaxed using DFT+$U$ with $U = 3.5$eV. Notably, the value of $U$ in the relaxation calculation do not affect the atomic positions in the converged structure. The calculated band structure with $U = 3.5$ eV and $U = 1$ eV are shown in Fig. S1a and b, respectively.

The γ (bonding $d_{z2}$) bands in the DFT+$U$ ($U = 3.5$eV) calculated band structure with experimental lattice parameters, as shown in Fig S1a, are located at ~320 meV below the Fermi level. On the other hand, the γ (bonding $d_{z2}$) bands in the DFT+$U$ ($U = 3.5$eV) calculated band structure for the unstrained case are located at ~110 meV below the Fermi level[2]. Thus, first principles calculation shows that with 2% in-plane compressive strain, the γ bands shift downwards by ~ 210 meV, compared to the unstrained case.

In the case of a 1% expansion of the $c$ lattice parameter at $U = 3.5$ eV, the effective pressure calculated from DFT+$U$ is 5.52 GPa. However, unlike the hydrostatic high-pressure case, the effective pressure under this strain is anisotropic: the in-plane pressures are 7.76 GPa and 7.03 GPa along the $a$ and $b$ directions, respectively, while the out-of-plane pressure is only 1.78 GPa.

For all the DFT calculations, we used projector augmented-wave (PAW) pseudopotentials, treating the f-electrons of the rare-earth elements as core states, and employed the generalized gradient approximation (GGA) with the Perdew-Burke-Ernzerhof (PBE) exchange-correlation functional[3]. To account for the electronic correlations, the Dudarev formulation of the on-site Hubbard $U$ correction was applied to all Ni sites[4]. The convergence criterion for ionic relaxation (EDIFFG) was set to $8 \times 10^{-5}$ eV for $La_3Ni_2O_7$. The electronic energy convergence threshold (EDIFF) was specified as $8 \times 10^{-6}$ eV. The Methfessel-Paxton smearing method, with a smearing width of 0.2 eV, was employed to account for metallic electronic states. A plane-wave kinetic energy cutoff of 520 eV was used for the basis set. The Brillouin zone was sampled using a Monkhorst-Pack grid of 8×8×2[5].

**Lattice structure indicated from $k_z$-dispersion**

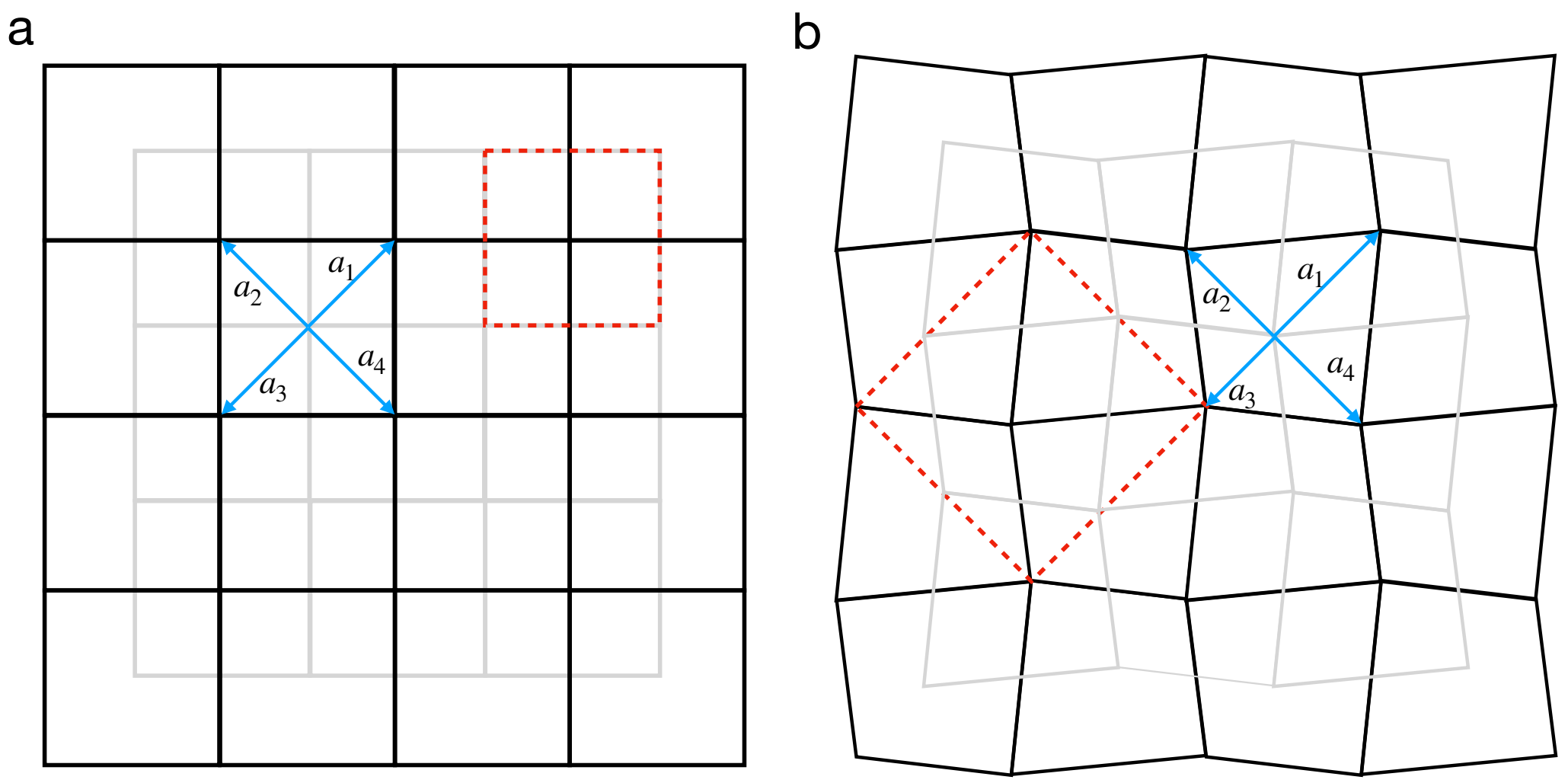

**FIG. S2.** two different lattice structures for a single Ni layer. (a) high symmetry case features a square lattice for the Ni-sites, where the unit cell can be chosen as the red dashed square. (b) low symmetry case (exaggerated) with the lattice sites indicating Ni. In this case, the unit cell is a 45-dgree rotated dashed square whose size is doubled compared to the case in (a). In both figures, the gray lattices indicate adjacent bilayer Ni lattice, and the four vectors $\vec{a}_i$ denote in-plane displacement of each Ni relative to its four nearest neighbor Ni in the other bilayer.

Here we discuss why and how the $k_z$-dispersion as measured in ARPES indicates a lower symmetry for the Ni sublattice. To this end, we first notice that the hopping terms between two adjacent bilayers, upon performing Fourier transform in the *x*-*y* plane, acquire some in-plane form factor that depends on the four vectors introduced in Fig. S2. For instance, if the Ni sites from each layer form a perfect square lattice, as shown in Fig. S2a, the inter-bilayer hopping of the $d_{x^2-y^2}$ orbitals, characterized by $t_\perp^{xx}$, is found to be proportional to

$$f(k_x, k_y) = e^{i\vec{k}\cdot\vec{a}_1} + e^{i\vec{k}\cdot\vec{a}_2} + e^{i\vec{k}\cdot\vec{a}_3} + e^{i\vec{k}\cdot\vec{a}_4},$$

where $\vec{k} = (k_x, k_y)$. If the four vectors $\vec{a}_i$ are as shown in Fig. S2a, the form factor reduces to $f(k_x, k_y) = cos(k_x a/2)cos(k_y a/2)$ where $a$ is the unit cell length. It is then straightforward to see that along the BZ boundary for $k_x, k_y$, the form factor $f(k_x, k_y)$ vanishes. This immediately implies that when either $k_x$ or $k_y$ is equal to $\pi/a$, such as at the M point, different bilayers effectively decouple due to the vanishing of $f(k_x, k_y)$. In this case, the $k_z$-dispersion cannot arise even if we take the apical oxygen sites within the Ni bilayer into account when constructing the tight-binding Hamiltonian.

There is another possibility though that can lead to finite $k_z$-dispersion at around M point when $f(k_x, k_y)$ vanishes, which is to assume the Wannier centers of $d_{x^2-y^2}$ orbitals and $d_{z^2}$ orbitals are shifted in *z*-direction. This will give rise to $k_z$ dependence of the dispersion from intra-bilayer hoppings when the inter-bilayer hoppings vanish. However, we do not observe Wannier center splitting from the DFT calculations[2].

Therefore, the only reason that ARPES data shows $k_z$-dispersion must be from the fact that the single Ni layer takes a lower symmetry case as shown in Fig. S2b. In this case, the unit cell is 2-times larger than the high symmetry case in Fig. S2a, and what's more important is that, although $f(k_x, k_y)$ is still given by the above equation, it does not vanish when $(k_x, k_y)$ is around M point. It is this fact that allows for finite $k_z$-dispersion. We note that Fig. S2b is also consistent with DFT calculations[2], although here we merely infer this structure by ruling out the high symmetry case, in order to be consistent with ARPES.

## Construction of a simple tight-binding model

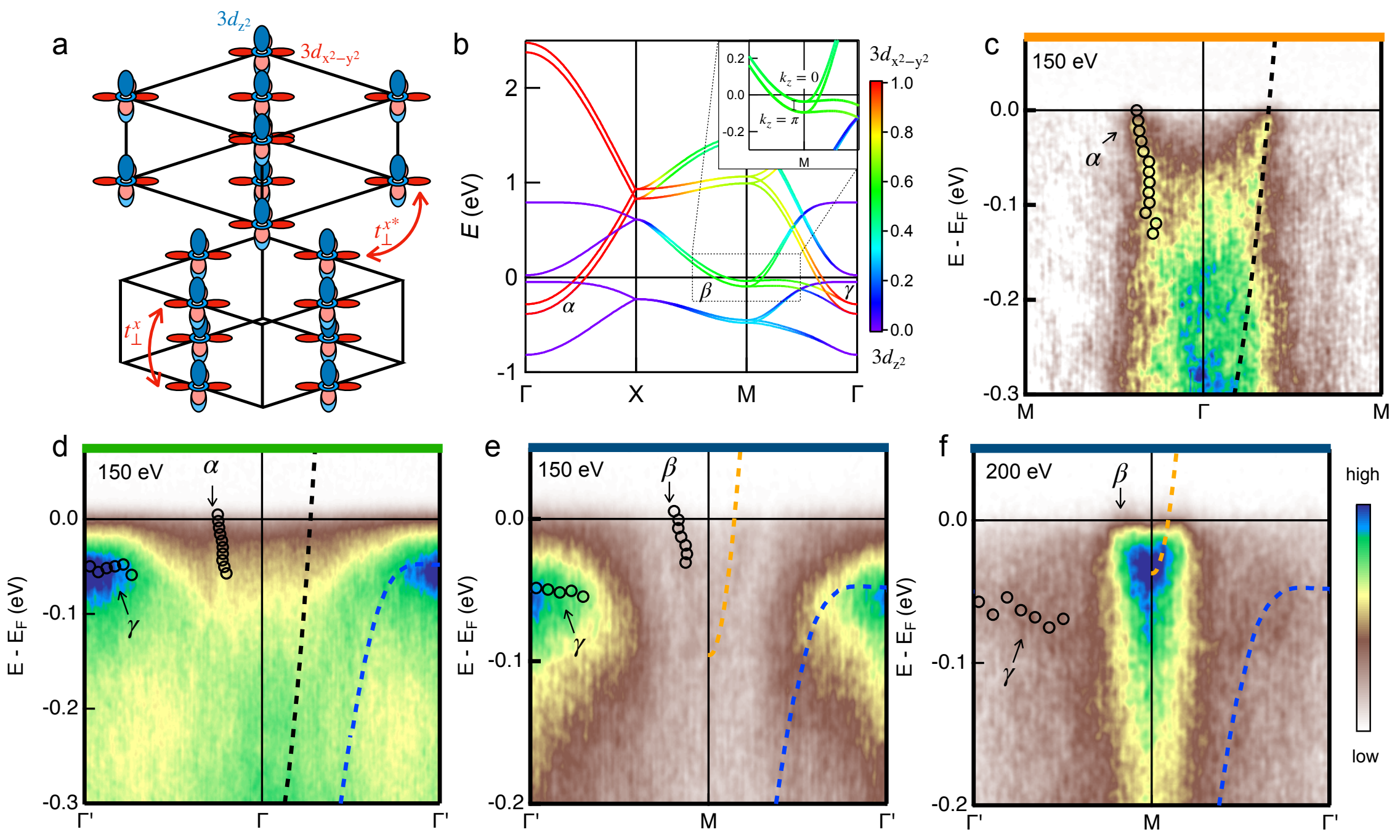

**FIG. S3.** Tight-binding description of the electronic structure. (a) Schematic of the bilayer LPNO lattice with the additional hopping parameter $t_\perp^{x*}$ illustrated. It corresponds to a hybridization between the $3d_{x^2-y^2}$ orbital (red) in the top NiO plane of the bottom bilayer and the same orbital in the bottom NiO plane of the top bilayer. The $3d_{z^2}$ orbital is colored in blue. (b) The tight-binding electronic structure based on parameters listed in Table II. The coloring describes the orbital projection weight and the inset provides a closer view of the β band $k_z$ dispersion. (c)-(f) Spectra along the three high-symmetry cuts of the sample presented in Fig. 2c-e, with the corresponding cut direction indicated by the frame colors (defined in Fig. 2a). The black circles mark the position of fitted Lorentzian peaks in MDCs describing the β band and the position of maximum intensity in EDCs describing the γ band. The black, orange, and blue dashed lines plot the dispersion of our tight-binding model for α, β, and γ bands, respectively.

To capture the observed $k_z$ dispersion, we construct a simple tight-binding model revised from existing literature[2]. The revised tight-binding Hamiltonian includes total of 16 orbitals, corresponding to 2 bilayers, with each bilayer adapting a low symmetry crystal structure containing 4 Ni. It is constructed as $H(k) = \begin{bmatrix} H_s(k) & H_z(k) \\ H_z^*(k) & H_s(k) \end{bmatrix}$, with $H_s(k) =$

$$\begin{bmatrix} H_A(k) & H_{AB}(k) \\ H_{AB}^*(k) & H_A(k) \end{bmatrix} \text{ and } H_z(k) = \begin{bmatrix} 0 & 0 & 0 & 0 & T_{\perp k}^{x*} & 0 & 0 & 0 \\ 0 & 0 & 0 & 0 & 0 & 0 & 0 & 0 \\ 0 & 0 & 0 & 0 & 0 & 0 & T_{\perp k}^{x*} & 0 \\ 0 & 0 & 0 & 0 & 0 & 0 & 0 & 0 \\ T_{\perp k}^{x*} & 0 & 0 & 0 & 0 & 0 & 0 & 0 \\ 0 & 0 & 0 & 0 & 0 & 0 & 0 & 0 \\ 0 & 0 & T_{\perp k}^{x*} & 0 & 0 & 0 & 0 & 0 \\ 0 & 0 & 0 & 0 & 0 & 0 & 0 & 0 \end{bmatrix}. \text{ Here, } H_A(k) =$$

$$H_B(k) = \begin{bmatrix} T_k^x & 0 & V_k^x & V_k^{xz} \\ 0 & T_k^z & V_k^{xz} & V_k^z \\ V_k^{x*} & V_k^{xz*} & T_k^x & 0 \\ V_k^{xz*} & V_k^{z*} & 0 & T_k^z \end{bmatrix}, H_{AB}(k) = \begin{bmatrix} t_\perp^x & 0 & 0 & V_{1\perp k}^{xz} \\ 0 & t_\perp^z & V_{1\perp k}^{xz} & 0 \\ 0 & V_{1\perp k}^{xz} & t_\perp^x & 0 \\ V_{1\perp k}^{xz} & 0 & 0 & t_\perp^z \end{bmatrix},$$

$$T_{\perp k}^{x*} = t_\perp^{x*} \cos(k_{x/y} r_\perp^x) \cos(k_z r_\perp^z)$$

The basis and definition of $T_k^x, T_k^z, V_k^x, V_k^{xz}, t_\perp^x, t_\perp^z, V_{1\perp k}^{xz}$ follow the description in ref.[2], except for $T_{\perp k}^{x*}$. It describes the additional inter-bilayer hopping illustrated in Fig. S3a, where $\overrightarrow{r_\perp}$ is the nearest neighbor displacement vector between the Ni site in the top layer of a bilayer and the Ni site in the bottom layer of the next bilayer above and $t_\perp^{x*}$ is the corresponding hopping parameter. The exact positions of each ion used in our tight binding model is taken from our DFT calculation described above, assuming a 2% biaxial in-plane compression and 1% *c*-axis expansion.

As shown in Fig. S3b, the introduction of a small inter-bilayer $3d_{x^2-y^2}$ to $3d_{x^2-y^2}$ hopping term produces a considerable $k_z$ dispersion of the β band without perturbing the γ band. Table SI lists the best set of tight-binding parameters that we find for describing the measured electronic structure. The corresponding tight-binding dispersions are plotted on top of the high-symmetry cuts in Fig. S3c-f in dashed lines with the black, orange, and blue colors corresponding to the α, β, and γ bands. We notice that it is difficult to capture all dispersions satisfactorily, with the measured dispersions considerably renormalized compared to tight-binding model predictions. With this caveat in mind, we point out that, compared to literature values for bulk samples, large changes concentrate in the parameters describing $3d_{z^2}$ orbitals. Both site energy and the in-plane next nearest neighbor hopping are enhanced while higher order as well as inter-layer hoppings get suppressed. As for the additional $t_\perp^{x*}$ hopping, its value is an order of magnitude smaller than in-plane nearest neighbor hopping and comparable to the in-plane second nearest neighbor hopping. Given this non-trivial magnitude, its effect becomes evident in the LPNO thin films.

| | | | | | |
|---|---|---|---|---|---|
| $t_1^x(t_{1'}^x)$ | $t_2^x(t_{2'}^x)$ | $\epsilon^x$ | $t_3^x$ | $t_\perp^x$ | $t_{1\perp}^{xz}$ |
| -0.345 | 0.057 | 0.851 | -0.008 | 0.0 | 0.01 |
| $t_1^z(t_{1'}^z)$ | $t_2^z$ | $\epsilon^z$ | $t_1^{xz}(t_{1'}^{xz})$ | $t_\perp^z$ | $t_\perp^{x*}$ |
| -0.096 | -0.051 | 0.192 | -0.18 | -0.42 | -0.051 |

**Table SI.** Tight-binding parameters of the revised 16 orbital model. The definition of the parameters follows table I in ref.[2], except for $t_\perp^{x*}$ which is defined in Fig. S3a.